\documentclass[aps,pre,print,onecolumn,tightenlines]{revtex4}
\usepackage{amsfonts}
\usepackage{amsmath}
\usepackage{amssymb}
\usepackage[colorlinks, citecolor=blue,linkcolor=red]{hyperref}
\usepackage{color}
\usepackage{float}
\usepackage{hyperref}
\usepackage{textcomp}
\usepackage{graphicx}
\usepackage{changes}

\begin{document}

\title{Stabilization of Axisymmetric Airy Beams by Means of Diffraction and Nonlinearity Management in Two-Dimensional Fractional Nonlinear Schr\"{o}dinger Equations}

\author{Pengfei Li}
\email{lipf@tynu.edu.cn}
\affiliation{Department of Physics, Taiyuan Normal University, Jinzhong, 030619, China}
\author{Yanzhu Wei}
\affiliation{College of computer science and technology, Taiyuan Normal University, 030619, China}
\author{Boris A. Malomed}
\affiliation{Department of Physical Electronics, School of Electrical Engineering, Faculty of Engineering, and Center for Light-Matter Interaction, Tel Aviv University, Tel Aviv 69978, Israel}
\affiliation{Instituto de Alta Investigaci\'{o}n, Universidad de Tarapac\'{a}, Casilla 7D, Arica, Chile}
\author{Dumitru Mihalache}
\affiliation{Horia Hulubei National Institute of Physics and Nuclear Engineering, Magurele, Bucharest RO-077125, Romania}

\begin{abstract}
The propagation dynamics of two-dimensional (2D) ring-Airy beams is studied
in the framework of the fractional Schr\"{o}dinger equation, which includes saturable or cubic
self-focusing or defocusing nonlinearity and L\'{e}vy index ((LI) alias for the fractionality) taking values $1\leq\alpha \leq 2$.
The model applies to light propagation in a chain of optical cavities emulating fractional diffraction.
Management is included by making the diffraction and/or nonlinearity coefficients periodic functions of
the propagation distance, $\zeta$. The management format with the nonlinearity coefficient decaying as $1/\zeta$ is
considered, too. These management schemes maintain stable propagation of the ring-Airy beams, which maintain their axial symmetry,
in contrast to the symmetry-breaking splitting instability of ring-shaped patterns in 2D Kerr media. The
instability driven by supercritical collapse at all values $\alpha < 2$ in the
presence of the self-focusing cubic term is eliminated, too, by the means of management.
\end{abstract}

\maketitle

\section{Introduction}

\label{Sec I}

The nonspreading solution of the Schr\"{o}dinger equation in the form of the
Airy function was found by Berry and Balazs in the context of quantum
mechanics in 1979 \cite{Berry-Balaz-1979AJP}. This solution follows a curved
parabolic trajectory, similar to a projectile moving under the action of
gravity. The prediction has been realized experimentally in the form of
Airy-shaped electron beams \cite{Arie}. Following the commonly known
principle that the propagation of classical optical beams in linear media
and the dynamics of a quantum particle are governed by essentially the same
Schr\"{o}dinger equation, nonspreading/nondiffracting packets, including
Airy waves, have been investigated in detail in linear optics \cite%
{Efremidis-2019-optica} and plasmonics \cite{plasmon}. Bessel beams are also
commonly known examples of diffraction-free waves, and were predicted and
experimentally demonstrated by Durnin {et al}. in 1987 \cite%
{Durnin-josaa-1987,Durnin-PRL1987}. Other classes of nondiffracting wave
modes, including Mathieu and parabolic beams, have also been investigated
\cite{Gutierrez-ol-2000,Bandres-ol-2000}; see, also,  review \cite{review}.

Airy beams have been created in optics by dint of inputs with properly
shaped intensity and phase \cite{Siviloglou-2007-PRL,Siviloglou-2007-OL}.
Those works demonstrated, in particular, self-healing of perturbed
beams. Cylindrically symmetric ring-Airy beams are able to abruptly
autofocus in the linear regime, thus delivering high-energy pulses into
transparent samples, as predicted theoretically \cite{Abruptly autofocus1}
and demonstrated experimentally \cite{Abruptly autofocus2,Abruptly
autofocus3}. By engineering the phase profile in the Fourier space, classes
of abruptly autofocusing Airy beams that follow different trajectories have also been
proposed \cite{Abruptly autofocus4,Abruptly autofocus5}. Propagation
dynamics of ring-Airy beams with embedded optical vorticity (angular
momentum) have also been investigated theoretically and experimentally \cite%
{ProRAB1,ProRAB2,ProRAB3}. Applications of Airy beams range from optical\mbox{
filamentation \cite{Filamentation1,Filamentation2,Filamentation3}}, imaging
\cite{Imaging1,Imaging2,Imaging3}, transportation and manipulation of
particles \cite{Particle Manipulation1,Particle Manipulation2,Particle
Manipulation3}, and to driving surface plasmon polaritons \cite{SPP1,SPP2,SPP3}.

The propagation of self-interacting high-power Airy beams has been studied
in the framework of Schr\"{o}dinger equations with various nonlinearities.
Accelerating self-trapped beams were thus predicted in Kerr, saturable,
and nonlocal nonlinear media \cite{Kaminer-PRL-2011,OE-19-23706}. For strong
Kerr nonlinearity, soliton shedding by Airy beams was analyzed in  \cite%
{OE-19-17298}. The latter phenomenon was also predicted in nonlocal
nonlinear media \cite{SciRep-5-9814}. Trajectories of Airy beams and pulses
have been studied experimentally in nonlinear photorefractive and Kerr media
\cite{OL-37-3201,OL-38-380}. Comparison of self-accelerating linear Airy
beams and solitons, which may feature similar dynamics in specific
two-component systems, was presented in a recent review \cite{EPL}.

The propagation of two-component Airy waves in second-harmonic-generating
media with quadratic nonlinearities was observed in \cite%
{Segev-Arie}. In this setting, it was further predicted that linear Airy
beams launched in the one- or two-dimensional (1D or 2D) second-harmonic
component generate sets of solitons through parametric instability\mbox{ \cite%
{Thaw1,Thaw2}}. An extension of the analysis was also developed for 1D and 2D
three-wave systems \cite{Thaw3,Thaw4}.

A non-Hermitian parity--time symmetric potential can modify the trajectory of
an Airy beam without affecting its ability for diffraction-free
propagation \cite{AnnPhy-1700307}. Specifically, Airy beams in a parity--time
symmetric Gaussian potential feature diffraction-free propagation over long
parabolic trajectories \cite{OC-410-717}.

Kinetic equations with partial fractional derivatives have been used to describe anomalous diffusion and relaxation phenomena,
including Hamiltonian chaos, disordered media, underground water pollution,
reactions in complex systems, and fractional diffusion in inhomogeneous media \cite{New1,New2,New3}.
The concept of fractional derivatives also appears in diverse areas of physical phenomenology,
such as the quantum Hall effect and the fractional Josephson effect \cite{New4,New5}.
Further, the fractional Schr\"{o}dinger equation (FSE) generalizes the classical Schr\"{o}dinger equation,
which is a canonical model for various physical phenomena, such as nonlinear optics,
hydrodynamics, and the Bose--Einstein condensates \cite{New6,New7,New8}.

Another vast research area in optics and related areas deals with the FSE.
It was introduced by Laskin \cite{FSE1,FSE2,FSE3}, who considered, by means
of the Feynman-integral technique, the evolution of the wave function of
quantum particles moving by L\'{e}vy flights (random jumps). The
aforementioned similarity between the quantum--mechanical linear Schr\"{o}%
dinger equation and the equation for the paraxial diffraction of light beams
suggests schemes for the realization of FSE\ in optics, as first proposed by
Longhi \cite{FSE4}. That work elaborated a scheme based on the so-called 4%
\textit{f} configuration in an optical cavity. The beam is transformed by a
lens from the coordinate space into the Fourier domain, in which a phase-shift emulating the fractional diffraction is introduced by means of an
appropriate phase plate. Then, the second lens casts the optical field back
into the spatial form, which carries the phase structure corresponding to
the fractional diffraction. Experimentally, this scheme was recently
realized in the temporal domain \cite{Shilong}.

The propagation of beams with various shapes under the action of
fractional paraxial diffraction has been addressed in the framework of FSE. In
particular, Airy beams have been studied in this context \cite%
{FSE5,FSE6,FSE7,FSE8,FSE9}. Furthermore, the self-interaction and
propagation dynamics of Airy beams have been investigated in the framework
of the fractional nonlinear Schr\"{o}dinger equation (NLSE), which adds the
Kerr term to the optical FSE \cite{FNLSE1,FNLSE2}; see, also, review \cite%
{review2}.

The above-mentioned possibility to implement fractional diffraction
implies that it is possible to build a system with an array of 4\textit{f}
setups with different parameters. This option, in turn, suggests one to
consider an FSE including a management scheme \cite{management}
that makes the effective diffraction coefficient a function of the
propagation distance, $z$. Moreover, the fact that the diffraction-emulating
phase shifts are introduced by the phase plates inserted into the optical
cavities makes it possible to emulate not only positive but also negative diffraction. The latter possibility allows one to define
management patterns in the form of periodic alternation of  artificial
diffraction between positive and negative strengths; see Equation (\ref{d(z)})
below. Such diffraction management schemes are akin to ones that have been
previously realized for normal (non-fractional) diffraction by  beam
propagation in waveguiding arrays with periodically varying orientation
of the guiding cores \cite{Silberberg}.

The subject of the present work is the dynamics of axially symmetric
ring-Airy beams governed by the 2D fractional NLSE with the aforementioned $%
z $-dependent local diffraction and/or nonlinearity coefficients. The paper
is organized as follows. The model is introduced in Section \ref{Sec II}, which
is followed by a detailed analysis of the dynamics of ring-Airy beams
in Section \ref{Sec III}. In particular, it is found that a saturable
self-focusing nonlinearity does not break the axial symmetry of the
ring-shaped beams through azimuthal instability. This is an essential
result, as similarly shaped modes are often vulnerable to that instability
\cite{Skryabin,AIP}. The paper is concluded in Section \ref{Sec IV}.

A more sophisticated management scheme may apply to the fractionality degree
(i.e., the L\'{e}vy index (LI)) in linear or nonlinear FSE, making LI a function of $z$; cf.  \cite{LImanagement}. Such a scheme
will be considered elsewhere.

\section{The Model}

\label{Sec II}

As outlined above, we consider beam propagation along the $z$-axis in a
2D nonlinear isotropic medium with saturable nonlinear correction to the
refractive index, which can be described by the fractional NLSE:
\begin{equation}
i\frac{\partial A}{\partial z}=\left[ \frac{1}{2k_{0}n_{0}}d\left( z\right)
\left( -\nabla _{\perp }^{2}\right) ^{\alpha /2}-k_{0}n_{\text{NL}}\right] A,
\label{FNLSE}
\end{equation}%
where $A(x,y,z)$ is the amplitude of the optical field, and $k_{0}=2\pi
n_{0}/\lambda $ is the wavenumber corresponding to carrier wavelength $%
\lambda $ and linear refractive index $n_{0}$. The fractional-diffraction
operator in Equation (\ref{FNLSE}), with LI, $\alpha $, belonging to interval $%
1\leq \alpha \leq 2$ and variable diffraction coefficient $d(z)$, is defined
as the 2D version of the Riesz derivative \cite{book,Zhang-CMA-2257}:%
\begin{equation}
\left( -\nabla _{\perp }^{2}\right) ^{\alpha /2}A\left( x,y\right) =\mathcal{%
F}^{-1}\left[ \left( k_{x}^{2}+k_{y}^{2}\right) ^{\alpha /2}\mathcal{F}A(x,y)%
\right] ,  \label{operator}
\end{equation}%
where $\mathcal{F}$ and $\mathcal{F}^{-1}$ are
2D operators of the direct and inverse Fourier transform, and $k_{x,y}$ are
wavenumbers conjugate to transverse coordinates $\left( x,y\right) $. The
nonlinear correction to the refractive index is represented by the term
\begin{equation}
n_{\text{NL}}=n_{2}\left\vert A\right\vert ^{2}/(1+s\left\vert A\right\vert
^{2}),  \label{nNL}
\end{equation}%
which features saturation with strength $s>0$. The saturable nonlinearity is
well-known to occur, in particular, in semiconductor-doped glasses and
photorefractive media \cite{Sat-r1,Sat-r2}. The saturation in this model is
a crucially important factor because, as is well-known, cubic
(unsaturated) self-focusing in the 2D fractional NLSE leads to supercritical collapse at all values $\alpha <2$ \cite{review2} (and to the
usual critical collapse in the case of non-fractional diffraction, $\alpha
=2 $); hence, the solitons are unstable in the absence of the saturation.
Note that, in the absence of the saturation and management ($s=0$, $d(z)=%
\mathrm{const}$), the collapse is driven by the symmetry (invariance) of Equation
(\ref{FNLSE}) with respect to the scaling transform:%
\begin{equation}
z=z_{0}\tilde{z},\left( x,y\right) =x_{0}\left( \tilde{x},\tilde{y}\right)
,A=A_{0}\tilde{A},z_{0}=x_{0}^{\alpha }=A_{0}^{-2}.  \label{scaling}
\end{equation}%
The saturation stabilizes the model against collapse by breaking the
scaling~symmetry.

Following the pattern of Equation (\ref{scaling}) but in the presence of the
saturation and diffraction management, the variables in Equation (\ref{FNLSE}) can be
normalized by means of rescaling:
\begin{equation}
\psi \left( \xi ,\eta ,\zeta \right) =\sqrt{s}A\left( x,y,z\right) , \left(
\xi , \eta \right) =\left( x,y\right)/\rho_{0} , \zeta =z/L,  \label{psi}
\end{equation}
where $\rho _{0}$ is the characteristic width of the input beam, and $%
L=k_{0}n_{0}d_{0}^{-1}\rho _{0}^{\alpha }$ is the respective diffraction
(Rayleigh) length corresponding to a characteristic value $d_{0}$ of the
diffraction coefficient. The accordingly scaled form of Equation (\ref{FNLSE}) is

\begin{equation}
i\frac{\partial \psi }{\partial \zeta }-\frac{1}{2}D\left( \zeta \right)
\left( -\nabla _{\perp }^{2}\right) ^{\alpha /2}\psi +\frac{\sigma
_{0}\left\vert \psi \right\vert ^{2}\psi }{1+\left\vert \psi \right\vert ^{2}%
}=0,  \label{FNLSE1}
\end{equation}%
where $D\left( \zeta \right) \equiv d(z)/d_{0}$ is the normalized
diffraction-modulation coefficient, and
\begin{equation}
\sigma _{0}\equiv \frac{k_{0}^{2}n_{0}n_{2}\rho _{0}^{\alpha }}{sd_{0}}
\label{sigma0}
\end{equation}%
with positive or negative values corresponds, respectively, to the
self-focusing or defocusing nonlinearity.

In particular, it is relevant to consider the case of the periodic
management corresponding to%
\begin{equation}
D(\zeta )=D_{0}\cos \left( \Omega \zeta \right)  \label{d(z)}
\end{equation}%
in Equation (\ref{FNLSE1}). As suggested by previous works on  dispersion and
diffraction management \cite{management}, a more general form of this format
may be taken as $D(\zeta )=\bar{D}+D_{0}\cos \left( \Omega \zeta \right) $,
including the path-average term, $\bar{D}$, which may be negative or
positive. Actually, simulations demonstrate that the latter term readily
becomes a dominant one in the ensuing evolution, making it similar to the standard model, with $D=\mathrm{const}$; therefore, here we
concentrate on the format (\ref{d(z)}), which produces most interesting
findings. By means of additional rescaling admitted by Equation (\ref{FNLSE1}),
we fix the modulation amplitude in Equation (\ref{d(z)}) as $D_{0}=1$.

To investigate the propagation characteristics of beams under the action
of different management formats, we also address decaying
diffraction-coefficient modulation:%
\begin{equation}
D\left( \zeta \right) =1/\zeta .  \label{1/zeta}
\end{equation}%
A similar modulation format has been considered in the case of  dispersion
management in nonlinear fiber optics \cite{Rome}.

Note that the substitution of%
\begin{equation}
\mathcal{Z}(\zeta )=\int D(\zeta )d\zeta  \label{Zzeta}
\end{equation}%
transforms Equation (\ref{FNLSE1}) into a fractional NLSE with nonlinearity
management:
\begin{gather}
i\frac{\partial \psi }{\partial \mathcal{Z}}-\frac{1}{2}\left( -\nabla
_{\perp }^{2}\right) ^{\alpha /2}\psi +\frac{\sigma (\mathcal{Z})\left\vert
\psi \right\vert ^{2}\psi }{1+\left\vert \psi \right\vert ^{2}}=0,
\label{nonlin} \\
\sigma (\mathcal{Z})\equiv \frac{\sigma _{0}}{D(\zeta (\mathcal{Z}))},
\label{sigma}
\end{gather}%
where $\zeta (Z)$ is a function inverse to $Z(\zeta )$. In particular, it is
seen that the management is eliminated by substitution (\ref{Zzeta}) in the
linear version of Equation (\ref{FNLSE1}), with $\sigma _{0}=0$. The model with
nonlinearity management is considered below, too, neglecting the
saturation (dropping the denominator in the nonlinear term) in Equation (\ref%
{nonlin}). The aim is to check the possibility of stabilization of the 2D
solitons by means of nonlinearity management against the above-mentioned
supercritical collapse driven by the unsaturated nonlinearity. This
possibility is suggested by the previously discovered mechanism of the
stabilization of fundamental solitons (rather than Airy beams) by
nonlinearity management against the onset of the critical collapse in the
case of $\alpha =2$ (the usual non-fractional diffraction)\mbox{ \cite%
{nonlin-management1,nonlin-management2,nonlin-management3,nonlin-management4}}%
.

As mentioned above, another version of the management, which is specific to
the fractional setting, can be introduced by periodic modulation of LI: $%
\alpha =\alpha (z)$. Unlike what is considered above, this form of the
management (that will be addressed elsewhere) cannot be eliminated from the
linearized equation.

\section{Numerical Results}

\label{Sec III}

\subsection{The Model with Diffraction Management}

In this section, we address the propagation of the ring-Airy beam governed
by Equation (\ref{FNLSE1}) with input%
\begin{equation}
\psi \left( \xi ,\eta \right) =\Psi _{0}\text{Ai}\left( \frac{r_{0}-r}{w_{0}}%
\right) \exp \left[ a\left( \frac{r_{0}-r}{w_{0}}\right) \right] ,
\label{Input1}
\end{equation}%
where Ai is the Airy function, $\Psi _{0}$ is the input's amplitude, $r=%
\sqrt{\xi ^{2}+\eta ^{2}}$ is the radial coordinate, $r_{0}$ and $w_{0}$
determine the initial radius and width, respectively, of the ring-Airy beam, and $a>0$ is
the exponential truncation factor that is necessary to secure convergence of
the integral power,%
\begin{equation}
P=\int \int \left\vert \psi \left( \xi ,\eta \right) \right\vert ^{2}d\xi
d\eta ,  \label{P}
\end{equation}%
cf. \cite{Siviloglou-2007-OL}.

Characteristic results can be presented using input (\ref{Input1}) with
\begin{equation}
A_{0}=1,r_{0}=10,w_{0}=1,a=0.1.  \label{input}
\end{equation}%

\begin{figure}[tbp]
\centering\includegraphics[width=12cm]{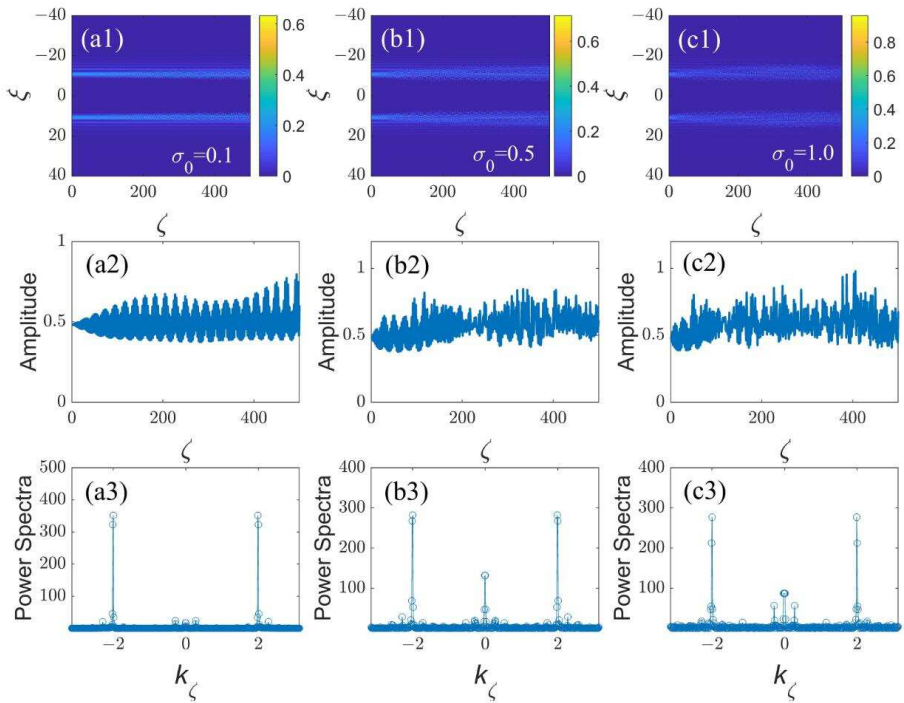}
\caption{The propagation dynamics of the ring-Airy beams under the combined
action of the self-focusing saturable nonlinearity, defined by Equation (\protect
\ref{nNL}) with values of $\protect\sigma _{0}$ indicated in the top panels,
and the periodic modulation of the diffraction coefficient, defined by Equation (%
\protect\ref{d(z)}) with $D_{0}=1$ and $\Omega =2$, in the framework of Equation (%
\protect\ref{FNLSE1}) with $\protect\alpha =1.5$. The propagation is
initiated by input (\protect\ref{Input1}) with parameters (\protect\ref%
{input}). (\textbf{a1},\textbf{b1},\textbf{c1}): Side views (i.e., the cross-section drawn through $%
\protect\eta =0$) of the propagating ring-Airy beams for different values of
nonlinearity parameter $\protect\sigma _{0}$ as indicated in the panels.
Panels (\textbf{a2},\textbf{b2},\textbf{c2}) show the amplitude of the optical field as the function of
the propagation distance. Panels (\textbf{a3},\textbf{b3},\textbf{c3}) present power spectra of the
amplitude from (\textbf{a2},\textbf{b2},\textbf{c2}): $k_{\protect\zeta }$ is the respective
wavenumber of the Fourier transform applied to the functions of $\protect%
\zeta $.}
\label{figure1}
\end{figure}

\begin{figure}[tbp]
\centering\includegraphics[width=12cm]{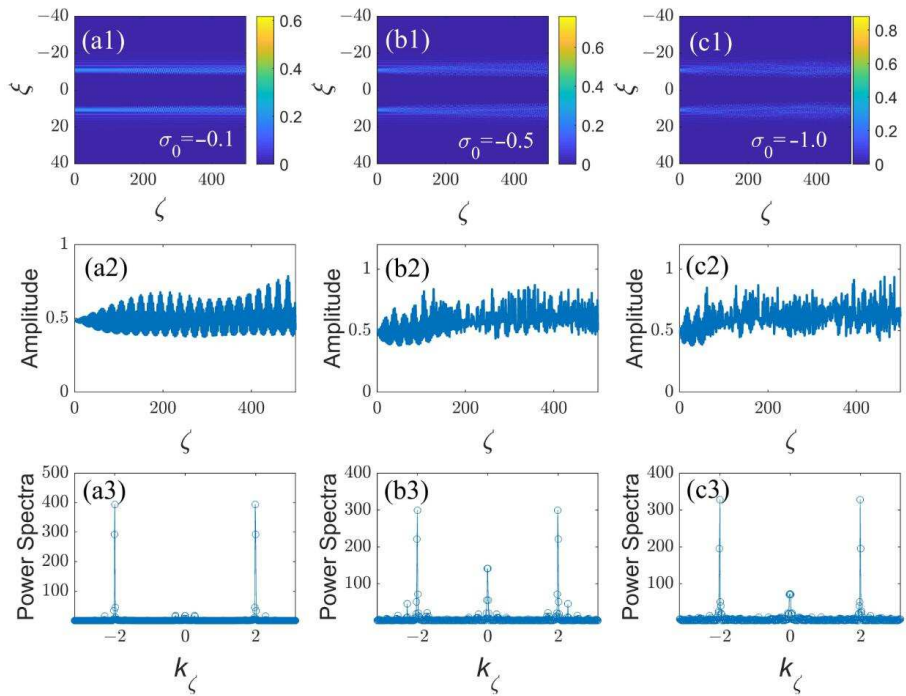}
\caption{The same as Figure \protect\ref{figure1} but for the case of
self-defocusing saturable nonlinearity defined by Equation (\protect\ref{FNLSE1}%
) with negative values $\protect\sigma _{0}$ indicated in the panels.(\textbf{a1},\textbf{b1},\textbf{c1}): Side views (i.e., the cross-section drawn through $%
\protect\eta =0$) of the propagating ring-Airy beams for different values of
nonlinearity parameter $\protect\sigma _{0}$ as indicated in the panels.
Panels (\textbf{a2},\textbf{b2},\textbf{c2}) show the amplitude of the optical field as the function of
the propagation distance. Panels (\textbf{a3},\textbf{b3},\textbf{c3}) present power spectra of the
amplitude from (\textbf{a2},\textbf{b2},\textbf{c2}): $k_{\protect\zeta }$ is the respective
wavenumber of the Fourier transform applied to the functions of $\protect%
\zeta $.}
\label{figure2}
\end{figure}

They are displayed in Figures \ref{figure1} and \ref{figure2} for the
self-focusing and self-defocusing nonlinearity, respectively.

Figure \ref{figure1}(a1,b1,c1) present the propagation dynamics with
different values of the\ self-focusing nonlinearity strength $\sigma _{0}$
in Equation (\ref{nNL}) and a fixed value of the modulation frequency, $\Omega =2$
in Equation (\ref{d(z)}) (recall $D_{0}\equiv 1$ is fixed by rescaling). In Figure %
\ref{figure1}(a1), the main lobe of the ring-Airy beam exhibits nearly
periodic oscillations in the presence of the weak nonlinearity. It is seen
in Figure \ref{figure1}(b1,c1) that the oscillations gradually extend
to side lobes of the beam\ as the strength $\sigma _{0}$ of the self-focusing
nonlinearity increases. The oscillation amplitude also gradually grows and
exhibits a trend towards chaotization with the increase of $\sigma _{0}$;
see Figure \ref{figure1}(a2,b2,c2). The respective power spectra of the
variable amplitude of the beam are displayed in Figure \ref{figure1}%
(a3,b3,c3). It is seen that the dominant spectral components, located at $%
k_{\zeta }=\pm 2$, are obviously determined by the aforementioned management
frequency ($\Omega =2$), while the apparent chaotization is produced, with
the growth of the nonlinearity strength, by the appearance of long-wave
spectral components around $k_{\zeta }=0$.

The results for the self-defocusing nonlinearity, which are displayed in
Figure \ref{figure2}, are generally similar. However, at stronger
nonlinearity, $\left\vert \sigma _{0}\right\vert =1$, the long-wave
component of the spectrum, which accounts for the chaotization, is, quite
naturally, broader (hence, more conspicuous) for the self-focusing sign; cf.
Figures \ref{figure1}(c3) and \ref{figure2}(c3).

\begin{figure}[tbp]
\centering\includegraphics[width=10cm]{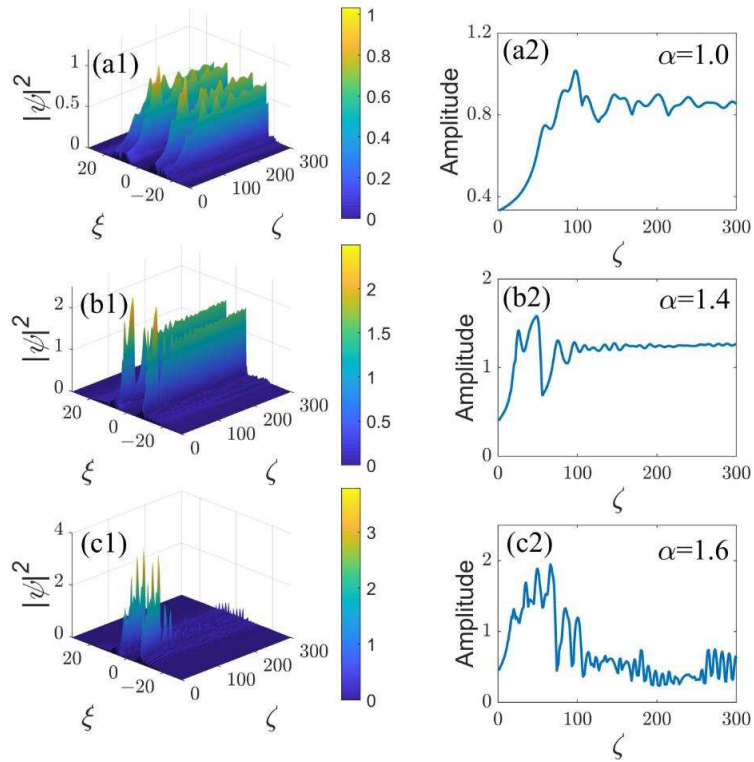}
\caption{The propagation dynamics of ring-Airy beams under the action of
the modulation format (\protect\ref{1/zeta}), as produced by simulations of
Equation (\protect\ref{FNLSE1}). Panels (\textbf{a1},\textbf{b1},\textbf{c1}) display, in the cross-section $%
\protect\eta =0$, the evolution of the propagating beams with the
self-focusing nonlinearity, $\protect\sigma _{0}=+1$, and LI values $\protect%
\alpha =1.0$, $\protect\alpha =1.4$, and $\protect\alpha =1.6$, as indicated
in the panels. The respective evolution of the amplitude of the optical
field is displayed in panels (\textbf{a2},\textbf{b2},\textbf{c2}).}
\label{figure3}
\end{figure}

Next, we address the propagation dynamics of the nonlinear beams under the
action of the decaying modulation format, defined as per Equation (\ref{1/zeta}).
Figure \ref{figure3}(a1,b1,c1) show that, in this case, the ring-Airy beams
naturally shrink, while their amplitude increases under the action of
self-focusing, following the decay of the diffraction strength. In the cases
of $\alpha =1.0$ and $\alpha =1.4$ (Figure \ref{figure3}(a2,b2)), the
increase in the amplitude is arrested and then it falls to a somewhat lower,
nearly constant value, which is conspicuously larger than the initial one,
due to the saturable character of the nonlinearity in Equation (\ref{nonlin}).
The situation is different in the case of $\alpha =1.6$, shown in Figure \ref%
{figure3}(c2), where the amplitude initially attains a higher value in the
course of the shrinkage, but then, due to a stronger effect of the
saturation, it falls back to a quasi-constant level, which is close to the
initial value.

\begin{figure}[tbp]
\centering\includegraphics[width=10cm]{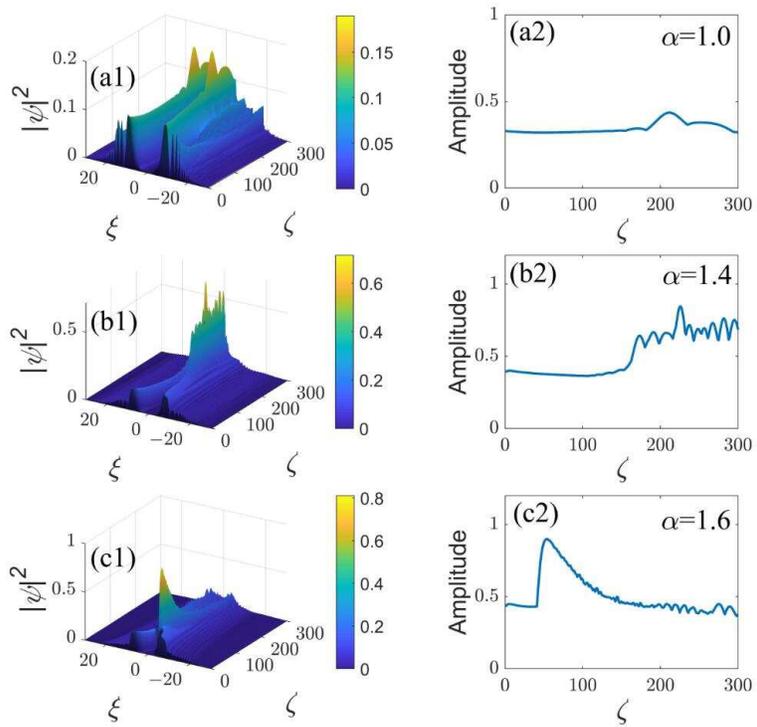}
\caption{The same as in Figure \protect\ref{figure3} but for the case of the
self-defocusing nonlinearity, with $\protect\sigma _{0}=-1$ in Equation (\protect
\ref{FNLSE1}).}
\label{figure4}
\end{figure}

In the case of the combination of the modulation format (\ref{1/zeta}) and
self-defocusing, corresponding to $\sigma _{0}=-1$ in Equation (\ref{FNLSE1}),
Figure \ref{figure4}(a1,b1,c1) show that the main lobe of the ring-Airy beams
tends to autofocus at a certain point, while the sidelobes do not
significantly shrink. In the cases of $\alpha =1.0$, relatively weak
autofocusing occurs at \mbox{$\zeta \approx 210$}, as seen in Figure \ref{figure4}%
(a1,a2). As the L\'{e}vy index increases, stronger autofocusing occurs
at shorter transmission distances in the cases of $\alpha =1.4$ and $\alpha
=1.6$ (\mbox{Figure \ref{figure3}(b1,c1)}, respectively).

\subsection{The Model with Nonlinearity Management}

Next, we consider the possibility of stabilization of ring-Airy beams by
means of nonlinearity management, modeled by the following fractional
NLSE:
\begin{equation}
i\frac{\partial \psi }{\partial \zeta }-\frac{1}{2}D(\zeta )\left( -\nabla
_{\perp }^{2}\right) ^{\alpha /2}\psi +\sigma \left( \zeta \right)
\left\vert \psi \right\vert ^{2}\psi =0,  \label{FNLSE2}
\end{equation}%
cf. Equation (\ref{nonlin}). Here, following \cite%
{nonlin-management1,nonlin-management2,nonlin-management3,nonlin-management4}%
, the spatially periodic management format is considered, with wavenumber $k$%
, \textit{viz}.:

\begin{equation}
\sigma (\zeta )=\sigma _{0}-\sigma _{1}\sin \left( \tilde{\Omega}\zeta
\right),  \label{nonlinear-mana}
\end{equation}%
cf. Equation (\ref{d(z)}), where the coefficient in front of $\sin \left( \tilde{%
\Omega}\zeta \right) $ is set equal so that $\sigma _{1}=1$ by means of
rescaling.

\begin{figure}[tbp]
\centering\includegraphics[width=12cm]{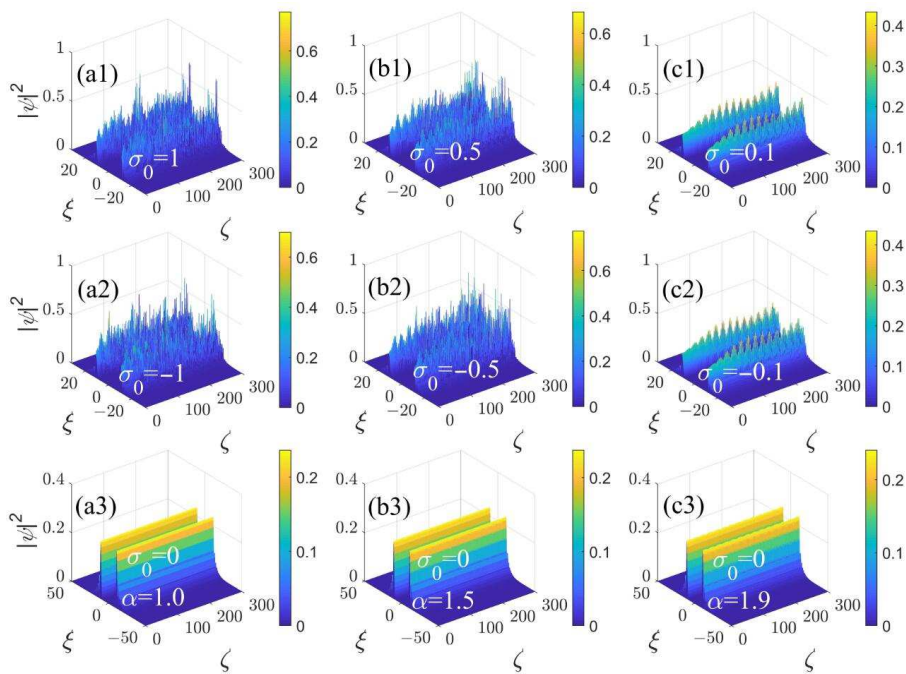}
\caption{The simulated propagation of the ring-Airy beams under the action
of diffraction management with modulation format (\protect\ref{d(z)}),
where $\Omega =2$ is set again in combination with nonlinearity
management (\protect\ref{nonlinear-mana}) with $\tilde{\Omega}=4$, as
produced by simulations of Equation (\protect\ref{FNLSE2}). Panels (\textbf{a1},\textbf{b1},\textbf{c1}) and
(\textbf{a2},\textbf{b2},\textbf{c2}) display, in the cross-section $\protect\eta =0$, the propagation
dynamics of the propagating beams with a fixed value of $\protect\alpha =1.5$
for different values of the self-focusing and defocusing nonlinearity
parameter $\protect\sigma _{0}$, as indicated in the panels. Panels
(\textbf{a3},\textbf{b3},\textbf{c3}) display the results for fixed $\protect\sigma _{0}=0$ and
different values of the LI: $\protect\alpha =1.0$, $\protect\alpha =1.5$,
and $\protect\alpha =1.9$, respectively. }
\label{figure5}
\end{figure}

The propagation dynamics of the ring-Airy beams are summarized in Figure \ref%
{figure5} for a fixed modulation wavenumber $\tilde{\Omega}=4$ under the
combined action of the periodic diffraction management, defined as per Equation (%
\ref{d(z)}), with $\Omega =2$, and nonlinearity management. In Figure \ref%
{figure5}(a,b1,c1,a2,b2,c2), the ring-Airy beams with $\alpha =1.5$ are
stabilized as the absolute value of constant term $\sigma _{0}$ in Equation (\ref%
{nonlinear-mana}) decreases in the cases of the\ self-focusing ($\sigma
_{0}>0$) and defocusing ($\sigma _{0}<0$) nonlinearity. It is seen that
stabilization of the beam against the above-mentioned supercritical collapse
can be achieved when the average value $\left\vert \sigma _{0}\right\vert $
in the modulation profile (\ref{nonlinear-mana}) is smaller than the
modulation amplitude (which is equal to $1$ in Equation (\ref{nonlinear-mana})).
Because the supercritical collapse is driven by the constant term $\sigma
_{0}$, while the effective stabilization is provided by the oscillatory one,
the stabilization is clearly seen to be most efficient at $\sigma _{0}=0$
for different values of $\alpha $ in Figure \ref{figure5}(a3,b3,c3), carrying
over into an apparently chaotic regime (but still not a collapsing one) with
the increase of $\sigma _{0}$. Fully clean stabilization at $\sigma _{0}=0$
and different values of LI $\alpha $ is clearly observed in \mbox{Figure \ref%
{figure5}(a3,b3,c3).}

\begin{figure}[tbp]
\centering\includegraphics[width=10cm]{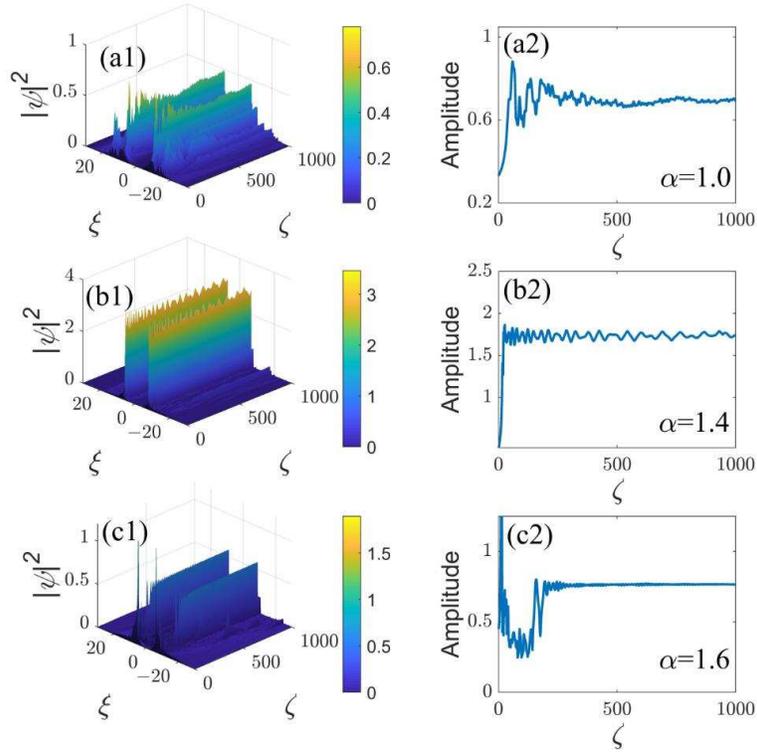}
\caption{The propagation dynamics of the ring-Airy beams under the combined
action of the decaying diffraction-modulation format (\protect\ref{1/zeta})
and nonlinearity management (\protect\ref{nonlinear-mana}) with $\protect%
\sigma _{0}=1$ and $\tilde{\Omega}=20$, as produced by simulations of Equation (%
\protect\ref{FNLSE2}). Panels (\textbf{a1},\textbf{b1},\textbf{c1}) display, in cross-section $\protect%
\eta =0$, the evolution of the beams with fixed values of $\protect\alpha $,
as indicated in the panels. Panels (\textbf{a2},\textbf{b2},\textbf{c2}) show the corresponding
evolution of the amplitude of the optical field.}
\label{figure6}
\end{figure}

As an additional test of the stabilization of the ring-Airy beams by the
nonlinearity management, the transmission of the beams was tested under the
combined effect of the fast management with $\tilde{\Omega}=20$ and $\sigma
_{0}=+1$ or $-1$, as shown with multiple examples in Figures \ref{figure6} and \ref%
{figure7}. It is observed that for different values of LI, the ring-Airy
beams autofocus after passing a certain distance, which is followed by
quasi-stabilization, with the amplitude converging to nearly constant values.

\begin{figure}[tbp]
\centering\includegraphics[width=10cm]{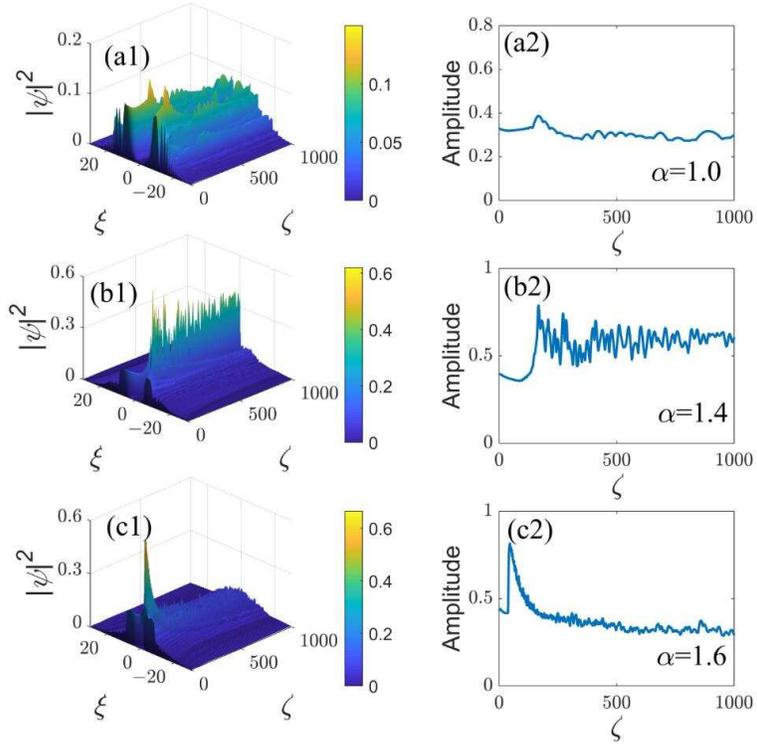}
\caption{The same as in Figure \protect\ref{figure6}, but for $\protect\sigma %
_{0}=-1$.}
\label{figure7}
\end{figure}

\section{Conclusions}

\label{Sec IV}

We have numerically investigated the propagation dynamics of 2D
ring-Airy beams with axial symmetry, governed by the fractional NLSE with
the saturable or cubic nonlinear term and variable diffraction and
nonlinearity coefficients. These equations, which belong to the general
class of management systems, model optical waveguides composed as an array
of elements emulating the fractional diffraction and carrying the Kerr
(saturable or non-saturable) nonlinearity. Adjusting the LI (L\'{e}vy index)
$\alpha $ and the modulation format of the fractional-diffraction
coefficient provides efficient means to maintain stable propagation of the
ring-Airy beams, keeping their symmetry unbroken. These findings are
essential, as normally, 2D ring-shaped patterns in Kerr media are unstable
against symmetry-breaking azimuthal perturbations that split the ring, and
moreover, all localized states are destroyed by the supercritical or
critical collapse at $\alpha <2$ or $\alpha =2$, respectively (the latter
value corresponds to the usual non-fractal diffraction). We have considered
the spatially periodic modulation profiles of the diffraction and
nonlinearity coefficients, as well as the diffraction-management profile
decaying inversely proportional to the propagation distance, in the case of
the saturable nonlinearity. The main lobe of the ring-Airy beam
exhibits nearly periodic oscillations under the action of periodic
modulation of the diffraction in the presence of the weak nonlinearity. The
oscillation amplitude gradually grows and exhibits a trend towards
chaotization with the growth of the nonlinearity strength. The ring-Airy
beams naturally shrink, while their amplitude increases under the action of
the decaying modulation format and self-focusing nonlinearity; then their amplitudes fall
to a somewhat lower, nearly quasi-constant value. In the case of the
combination of the decaying modulation and self-defocusing, the main lobe of
the ring-Airy beams tends to autofocus at a certain point, while the
sidelobes do not significantly shrink. Generally, the ring-Airy beams are
stabilized by means of nonlinearity management both for the periodic and
decaying modulation formats of diffraction. The management schemes
introduced in this work offer new possibilities for manipulating and
controlling the ring-shaped optical beams that would be unstable otherwise.
As an extension of the work, it is relevant to address the propagation of
the ring beams carrying angular momentum (vorticity).

\section{Acknowledgments}
This work was supported by the National Natural Science Foundation
of China (NSFC) (No. 11805141), the Applied Basic Research Program of Shanxi
Province (No. 201901D211424), and the Scientific and Technological Innovation
Programs of Higher Education Institutions in Shanxi (STIP) (2021L401). The
work of BAM was partly supported by the Israel Science Foundation through
grant No. 1695/22.

\end{document}